\documentstyle[11pt,Cargesepasp,twoside,epsf]{article}
\reversemarginpar

\begin{document}
\title{Wide-field (sub)millimeter continuum surveys of protoclusters: 
Clues to the origin of the IMF}
  \author{Fr\'ed\'erique Motte}
  \affil{California Institute of Technology, Mail Stop 320-47, 
1200 E. California Blvd., Pasadena, CA 91125, USA\\
Max-Planck-Institut f\"ur Radioastronomie, Auf dem H\"ugel 69, 
53121 Bonn, Germany} 
  \author{Philippe Andr\'e}
  \affil{CEA, DSM, DAPNIA, Service d'Astrophysique, C.E.~Saclay, 
F-91191~Gif-sur-Yvette~Cedex, France}

\begin{abstract}
Recent (sub)millimeter continuum surveys of nearby star-forming
regions have revealed a wealth of new, cold cloud fragments.  Those
which are small-scale (diameter~$ \la 10\,000$~AU), starless, and
gravitationally bound are good candidates for being the direct
progenitors of protostars, i.e., the structures within which
individual protostellar collapse is initiated.  The mass spectrum of
these protocluster condensations is reminiscent of the stellar initial
mass function (IMF), suggesting the IMF is partly determined by cloud
fragmentation at the pre-stellar stage of star formation.
\end{abstract}


\section{Introduction}

The question of the origin and possible universality of the initial
mass function (IMF), which is crucial for both star formation and
galactic evolution, remains a matter of debate (e.g. Elmegreen, this
volume). In the past, numerous molecular line studies of cloud
structure have attempted, without success, to relate the mass spectrum
of observed clumps to the stellar IMF (see, e.g., Williams, Blitz, \&
McKee 2000 and references therein). The reason of this failure is
presumably that most CO clumps are not gravitationally bound and
reflect more the characteristics of the low-density outer parts of
molecular clouds than the initial conditions of protostellar collapse
(e.g.\ Kramer et al.\ 1998).

\begin{figure}
\vskip -1.cm
\plotone{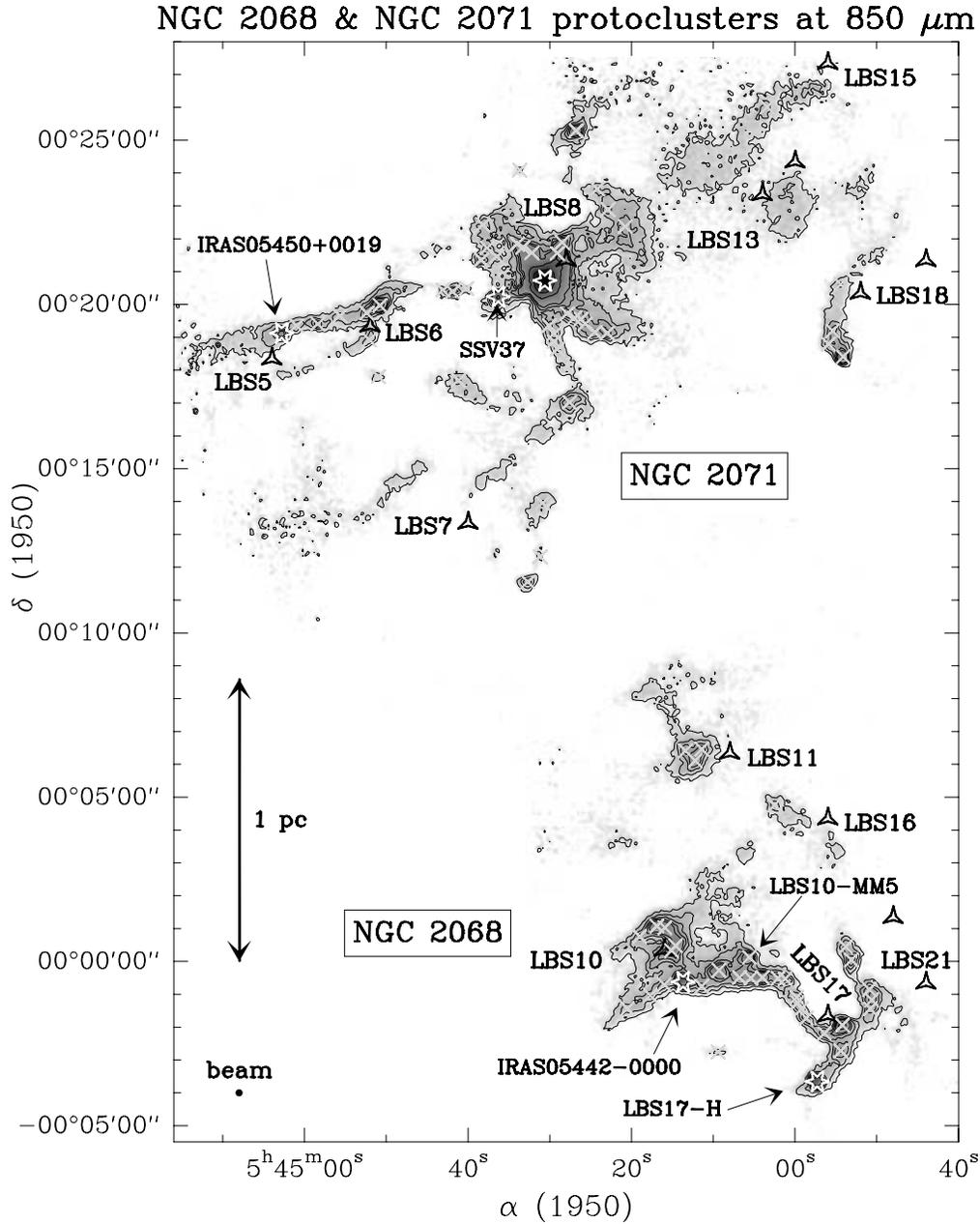}
\vskip -1.5cm
\caption{Dust continuum mosaic of the NGC~2068 and NGC~2071 
protoclusters at $850\:\mu$m, including the CS dense cores LBS5 to
LBS8, LBS10, LBS11 and LBS13 to LBS21 (from Lada, Bally, \& Stark
1991, marked with triangles). The data were taken with the SCUBA
bolometer array at the JCMT.  Starless condensations are denoted by
crosses and young embedded stars by star markers.}
\end{figure}

The advent of sensitive bolometer arrays on large (sub)mm
radiotelescopes has recently made possible extensive surveys for
protostars and their prestellar precursors in nearby star-forming
regions (see, e.g., review by Andr\'e, Ward-Thompson, \& Barsony
2000).  Wide-field millimeter continuum mapping of molecular clouds
was first performed using the (19- and 37-channel) MPIfR bolometer
arrays (MAMBO) at the IRAM~30m telescope (e.g. Chini et al. 1997;
Motte, Andr\'e, \& Neri 1998 -- hereafter MAN98).  More recently, the
bolometer cameras SCUBA and SHARC have been used at the JCMT and CSO
radiotelescopes (e.g. Lis et al. 1998; Johnstone et al. 2000 --
hereafter JWM00; Motte et al. 2001 -- hereafter MAWB01, see Fig.~1).
At the same time, mosaics of star-forming cloud cores have been
obtained with the Owens Valley millimeter array (OVRO), over somewhat
smaller areas but with higher angular resolution than single-dish
bolometer observations (e.g. Testi \& Sargent 1998 -- hereafter TS98).

These (sub)mm continuum surveys have detected a large number of
starless cloud fragments in the $\rho$~Oph (MAN98; JWM00), Serpens
(TS98), OMC1 (Coppin et al. 2000 -- hereafter CGJH00), NGC~1333
(Sandell \& Knee 2000 -- hereafter SK01), and NGC~2068/2071 (MAWB01)
cluster-forming regions.  Some of the starless fragments, dubbed here
{\it protocluster condensations}, are likely to be the direct
progenitors of protostars (see Sect.~2.2 below).  When such
condensations are carefully extracted from their surrounding cloud
(see Sect.~2.3), the associated mass spectrum is found to resemble the
shape of the stellar IMF (Sect.~3.1). This has important implications
for the origin of the IMF in clusters (Sect.~4).  We here give an
overview of these recent submillimeter results and compare the methods
employed in the various published studies.

\section{Analysis of cloud structure}

Relating cloud structure to star formation is a difficult task which
requires a careful analysis motivated by physical questions such as:
What is the size scale of protostellar collapse? How can we select
those cloud fragments that are the potential progenitors of
protostars? How can we accurately measure the mass reservoirs involved
in individual collapse? Possible answers are proposed below.

\subsection{Estimating the characteristic size scale of protostellar 
collapse}

Numerous CO studies have shown that the overall structure of molecular
clouds is fractal and probably shaped by turbulence, in both quiescent
and star-forming regions (e.g. Elmegreen \& Falgarone 1996; Heithausen
et al. 1998). On the other hand, there is a growing body of evidence
that this fractal structure breaks down once self-gravity takes over
at the high densities and small lengthscales characteristic of
individual protostellar collapse.\\
First, the CO maps of the Taurus cloud cannot be described by a single
fractal dimension (Blitz \& Williams 1997).  Second, the dense cores
observed within molecular clouds appear to be ``coherent'', i.e.,
largely devoid of turbulence, in their inner 0.1~pc$\ \simeq
20\,000$~AU parts (Goodman et al. 1998).  Third, mid-infrared
absorption images of isolated starless dense cores taken with ISOCAM
aboard ISO show that these cores are typically finite-sized structures
(Bacmann et al. 2000; Bacmann, Andr\'e, \& Ward-Thompson, this
volume).  Finally, (sub)mm continuum maps of optically thin dust
emission also indicate finite sizes for protostellar envelopes.  The
radial intensity profiles observed for the envelopes of nearby
protostars are indeed consistent with a power-law density structure
merging into some background cloud emission beyond some outer radius
$R_{\rm out} $ (MAN98; Motte \& Andr\'e 2001; see also Fig.~2a).  The
envelope outer radii are measured to be typically $\sim 5\,000$~AU in
$\rho$~Oph and NGC~2068/2071 while being $\ga 10\,000$~AU in Taurus.
These outer sizes may result from truncation by the local ambient
cloud pressure $P_{\rm s} $, varying from $<P_{\rm s}> \sim 5-10
\times 10^5 \ k_{\rm B}\ \mbox{cm}^{-3}\,$K in dense protoclusters
such as $\rho$~Oph and NGC~2068/2071, to $<P_{\rm s}> \sim 0.5\times
10^5 \ k_{\rm B}\ \mbox{cm}^{-3}\,$K in regions of more distributed
star formation such as Taurus.

{\em The typical size observed for (young) protostellar envelopes in a
given star-forming region provides a natural lengthscale which is
likely to be characteristic of the detached cloud fragments
participating in individual protostellar collapse.}

\begin{figure}
\plotfiddle{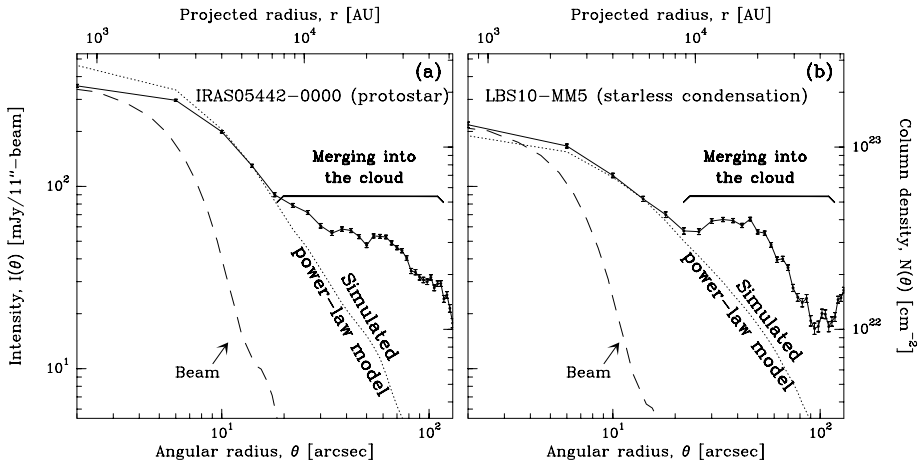}{9.5cm}{-90}{48.5}{48.5}{-200}{290}
\vspace{-3.2cm}
\caption{
Circularly averaged radial intensity profiles of a protostellar
envelope ({\bf a}) and a starless condensation ({\bf b}) in the
NGC~2068 protocluster, derived from a 1.3~mm continumm map taken with
MAMBO at the IRAM~30m telescope. Both profiles are clearly resolved
and merge into some background cloud emission at $R_{\rm out} \sim
6\,000-8\,000$~AU.  This behavior is qualitatively similar to that
observed in $\rho$~Oph (see Figs.~4c-d of MAN98), Taurus, and Perseus
(see, e.g., Figs.~3f and 4g of Motte \& Andr\'e 2001; Andr\'e, Motte,
\& Belloche, this volume).}
\end{figure}

\subsection{Selecting the possible progenitors of protostars}

Guided by the ideas outlined in \S~2.1 above, MAN98 and MAWB01 have
used a multiresolution wavelet technique (e.g.\ Starck, Murtagh, \&
Bijaoui 1998) to analyze their (sub)mm maps of the $\rho$~Oph and
NGC~2068/2071 protoclusters. With this technique, it is possible to separate
small-scale fragments from structures larger than the characteristic
outer size of protostellar envelopes. Note that the wavelet analysis 
is not subject to human bias, unlike the ``visual inspection method''
used by e.g. SK01 in NGC~1333.\\
As illustrated in Fig.~3 for the $\rho$~Oph-F region, the procedure
consists in decomposing the original image in a number of views of the
same field at different spatial scales.  For simplicity, only two such
views are shown in Fig.~3. The ``small-scale'' view (bottom right of
Fig.~3) has been adjusted so as to trace circumstellar envelopes (or
disks) around embedded young stellar objects (YSOs). In addition, the
same view reveals several compact starless fragments (e.g. F-MM1 and
F-MM2 in Fig.~3).  These are not associated with any centimeter radio
continuum emission or near-/mid-IR emission (MAN98). They are even
often seen in {\it absorption} by ISOCAM at 7~$\mu$m against the
diffuse mid-IR emission arising from the cloud's outer layers
(e.g. Motte et al. 1998b). These small-scale fragments are good
candidates for being pre-stellar condensations (see below).\\
On the other hand, the structures that show up in the ``large-scale''
view (bottom left of Fig.~3) correspond to dense cores already known
from DCO$^+$, CS, or NH$_3$ studies with $\sim\,$arcmin resolution
(e.g. Loren, Wootten, \& Wilking 1990).  In protoclusters, such dense
cores are typically the sites of multiple protostellar collapse and
correspond to the background environment of several smaller-scale
condensations. In regions of isolated star formation (e.g. Taurus),
dense cores of similar sizes generally give birth to only one or two
stars (cf. Myers 1998).

\begin{figure}
\plotfiddle{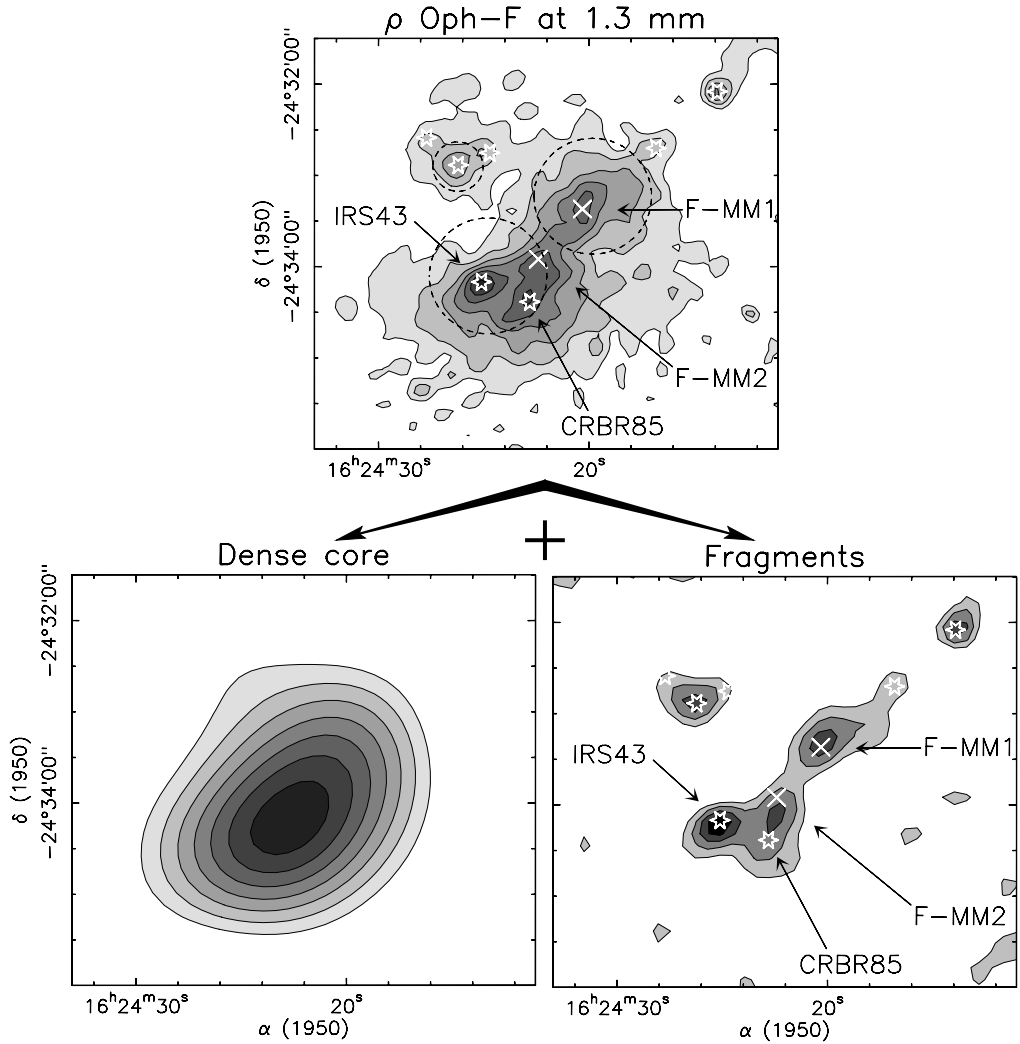}{15.5cm}{0}{75}{75}{-220}{0}
\vspace{-4.3cm}
\caption{Illustration (in the case of $\rho$~Oph-F) of the 
multiresolution wavelet technique used by MAN98 to analyze the
structure of the $\rho$~Oph cloud. The original 1.3~mm continuum map
(top) is decomposed into a large-scale view sensitive to structures
$\ga 10\,000$~AU in diameter (bottom left) and a small-scale view
tracing small-scale fragments $\sim 1\,000-6\,000$~AU in diameter
(bottom right).  The dashed large circles represent the sizes of the
clumps identified by JWM00.}
\end{figure}

The $\rho$~Oph starless condensations identified in this way by MAN98
have relatively narrow linewidths and are close to gravitational
virial equilibrium according to recent observations in high-density
molecular tracers (see Belloche, Andr\'e, \& Motte, this volume).
Furthermore these condensations, as well as those of MAWB01 in
NGC~2068/2071 have spheroidal shapes (aspect ratio $\sim 0.6$, see
also CGJH00). All of this suggests that the condensations have
dissipated (most of) their initial magnetic turbulent support and are
on the verge of collapse. In $\rho$~Oph, some of them already show
spectroscopic evidence of inward motions (see Belloche et al., this
volume).
Based on these characteristics and the similarity with protostellar
envelopes, we argue that the starless condensations identified in the
(sub)mm continuum are likely to be {\it pre-stellar in nature and the
direct progenitors of accreting (i.e. Class~0, Class~I) protostars}.

In the Serpens core, the OVRO interferometer study of TS98 filtered
out all cloud structures larger than $\sim 9\,000$~AU in diameter. As
this is comparable to the size of protostellar envelopes in Serpens,
the sources detected by TS98 are probably ``protocluster
condensations'' in the same sense as above. However, line and
centimeter radio continuum observations would be needed to confirm
that these condensations are self-gravitating and pre-stellar in
character.

In their independent study of the $\rho$~Oph protocluster, JWM00
analyzed cloud structure without reference to any circumstellar
lengthscale. In practice, they nevertheless selected fragments smaller
than $21\,000$~AU in diameter, owing to the Gaussian filter applied to
remove reduction artifacts from the SCUBA data.  They identified as
``clumps'' all the structures associated with a local maximum in their
map, regardless of size, global shape, or substructure.  Some of these
clumps are very diffuse and may be transient features confined by
external pressure rather than bound by gravity (cf. Bertoldi \& McKee
1992).  Moreover, several of the JWM00 clumps are and/or will clearly
be the sites of multiple protostellar collapse (e.g. the clump
associated with IRS43, CRBR85, and F-MM2 in Fig.~3).  Thus, in
general, the structures selected by JWM00 do not seem to be truly
representative of collapse initial conditions and cannot be taken as
the direct progenitors of protostars.

\subsection{Making reliable measurements of the collapse mass reservoir}

In order to estimate the total mass available for local gravitational
collapse, one first needs to quantify the importance of the background
cloud surrounding each pre-stellar condensation. This is definitely
the most difficult step in the analyses of MAN98, CGJH00, and MAWB01.
CGJH00 assumed that all their condensations were essentially unresolved
(i.e. diameter~$\la 6\,000$~AU in their case) and thus estimated only
lower limits to the corresponding mass reservoirs.  The analysis of
MAN98 and MAWB01 is more sophisticated as it uses the radial intensity
profiles of the condensations to determine their outer radii
(e.g. \S~2.1 and Fig.~2b). This makes it possible to filter the background 
very effectively and to get an accurate
measurement of the collapse mass reservoir, without any a~priori
assumption about the fragment shape.  It is however a time-consuming
process that cannot be easily automated.  The condensations must also
be spatially resolved which, in practice, limits the technique to
nearby regions.\\
As for TS98, they integrated the emission detected toward the
point-like sources identified in their OVRO 3.3~mm mosaic. They did
not correct for the physical size of the mass reservoir which may be
slightly larger than the $9\,000$~AU diameter of the interferometric
filter.\\
In another approach, JWM00 used a variant of the clump finding
algorithm of Williams, de~Geus, \& Blitz (1994) which assumes that a
fragment ends at the last significant contour surrounding it. Such a
boundary is signal-to-noise dependent and not necessarily related to
the physical outer radius resulting from, e.g., truncation by the
pressure of the background cloud.

After source extraction and background subtraction, the masses of the
condensations identified are estimated from the measured (sub)mm
fluxes assuming dust properties (temperature and opacity) adapted to
pre-stellar cores (see e.g. MAN98). At least far from massive YSOs,
starless condensations are expected to have low temperatures
$T_{\mbox{\tiny dust}} \la 10-15$~K, as they are well shielded from
external heating by cosmic rays and the interstellar radiation field
(e.g. Masunaga \& Inutsuka 2000; Evans et al. 2001).

\section{Overview of results}

With present instruments, wide-field (sub)mm surveys of nearby
($d<500$~pc) protoclusters are limited to the detection of $\ga
0.02~M_\odot$ condensations and are only complete down to $\ga
0.1~M_\odot$.  Table~1 lists several of the above-mentioned submm
continuum studies, along with the related molecular line survey of
Taurus by Onishi et al. (2001, this volume -- herafter OMKF01).  The
characteristics of the cloud fragments found in each study are given
in the last column.

\small
\begin{table}[htbp]
\caption{Summary of recent studies of cloud fragmentation}
\begin{tabular}{lll}
\tableline
Authors	\& region & Observations & Fragment characteristics$^{(1)}$\\
\tableline\tableline
~~~~MAN98		& At 1.3~mm with 	& 59, pre-stellar and self-gravitating \\
$\rho$~Oph, 150~pc	& MAMBO/IRAM~30m	& $1\,000-6\,000$~AU, $0.05-3~M_\odot$\\
\tableline 
~~~~TS98		& At 3~mm		& 26, pre-stellar\\
Serpens, 310~pc		& with OVRO		& $700-9\,000$~AU, $0.5-25~M_\odot$ \\ 
 \tableline
~~~~CGJH00		& At $850\ \mu$m with	& 39, pre-stellar or protostellar\\
OMC1, 450~pc		& SCUBA/JCMT		& $\la 6\,000$~AU, $0.1-100~M_\odot$ \\ 
\tableline
~~~~JWM00		& At $850\ \mu$m with	& 34, starless and 
still evolving\\
$\rho$~Oph, 150~pc	& SCUBA/JCMT		& $\sim 1\,500-10\,000$~AU, $0.03-10~M_\odot~^{(2)}$\\ 
\tableline
~~~~SK01		& At $850\ \mu$m with	& 33, pre-stellar or protostellar\\
NGC~1333, 220~pc	& SCUBA/JCMT		& $500-8\,000$~AU, $0.02-2.5~M_\odot~^{(2)}$ \\
\tableline
~~~~MAWB01		& At $850\ \mu$m with	& 70, pre-stellar\\
NGC2068/71, 400~pc	& SCUBA/JCMT		& $1\,600-13\,000$~AU, $0.3-6~M_\odot$ \\ 
\tableline 
~~~~OMKF01		& H$^{13}$CO$^+(1-0)$	& 45, starless\\
Taurus, 150~pc		& at the NRO 45~m 	& $<20\,000$~AU, $0.4-25~M_\odot$\\
\tableline\tableline
\end{tabular}
\begin{list}{}{}
\item[]{}$^{(1)}$~{Number, pre-stellar or protostellar nature, range 
of (deconvolved) FWHM sizes and masses.}
\item[]{}$^{(2)}$~{Values modified using $T_{\mbox{\tiny dust}}=15$~K 
instead of $T_{\mbox{\tiny dust}}=20-25$~K.}
\end{list}
\end{table}
\normalsize

\subsection{Mass distribution of protocluster condensations}

When one carefully selects the (sub)mm fragments seen on the same
spatial scales as protostellar envelopes (cf. \S~2.2), the resulting
mass spectrum appears to mimic the shape of the stellar IMF (see
MAN98; TS98; MAWB01 and Fig.~4). The cumulative mass spectra of the
pre-stellar condensations identified in the $\rho$~Oph, Serpens, and
NGC~2068/2071 protoclusters approximately follow the Salpeter
power-law (i.e. $N(>m)\propto m^{-1.35}$, Salpeter 1955) for
$m>0.3-0.5~M_\odot$, $\ga 0.4~M_\odot$ and $\ga 0.8~M_\odot$,
respectively.  At lower masses, the $\rho$~Oph mass spectrum flattens
to $N(>m)\propto m^{-0.5}$ as does the stellar IMF (e.g. Kroupa, Tout,
\& Gilmore 1993).  This suggests that {\em the protocluster
condensations detected in the (sub)mm continuum will form individual
stars/systems with a high efficiency roughly independent of mass}.
More quantitatively, comparison of the pre-stellar mass spectrum of
MAN98 with the mass spectrum determined by Bontemps et al. (2001) for
pre-main sequence objects suggests that the local star formation
efficiency is $\ga ~50-70\%$ for each prestellar condensation of
$\rho$ Oph.\\
Interestingly, the starless H$^{13}$CO$^+$ condensations detected by
Onishi et al. in Taurus also have a mass spectrum that resembles, for
$m>3.5~M_\odot$, the IMF of field stars (OMFK01, this volume).  In
marked contrast with CO results (see Fig.~4 and, e.g., Williams et
al. 2000), this finding suggests that high density molecular tracers
can also give access to the real progenitors of protostars.  The star
formation efficiency estimated by OMKF01 for their condensations is
significantly lower ($\la 15\%$) than that inferred in protoclusters,
which may be characteristic of regions forming stars in relative
isolation such as Taurus.

On the other hand, the mass distribution found by submm continuum
studies which do not carefully select small-scale pre-stellar
fragments (see Sect.~2.2) is generally scale-free and tends to follow
the same power-law as that observed for CO clumps (e.g. CGJH00 and
SK01). Likewise, the mass spectrum obtained by MAN98 when combining
the dense cores {\it and} compact condensations of $\rho$~Oph together
is roughly scale-free with $N(>m)\propto m^{-0.5}$.\\
The introduction of an upper sizescale in the study of cloud structure
naturally creates a break in this self-similar scaling.  This may
explain why JWM00, who filtered dust structures larger than
$21\,000$~AU in diameter in $\rho$~Oph, found a mass spectrum
reminiscent of the Salpeter IMF above a break at $m \sim 0.9~M_\odot$
(when $T_{\mbox{\tiny dust}}$=15~K).  However, the upper sizescale
used to select protocluster condensations must be physically motivated
(cf. \S~2.1) for the derived mass spectrum to be meaningful.

\begin{figure}[hbtp]
\plotone{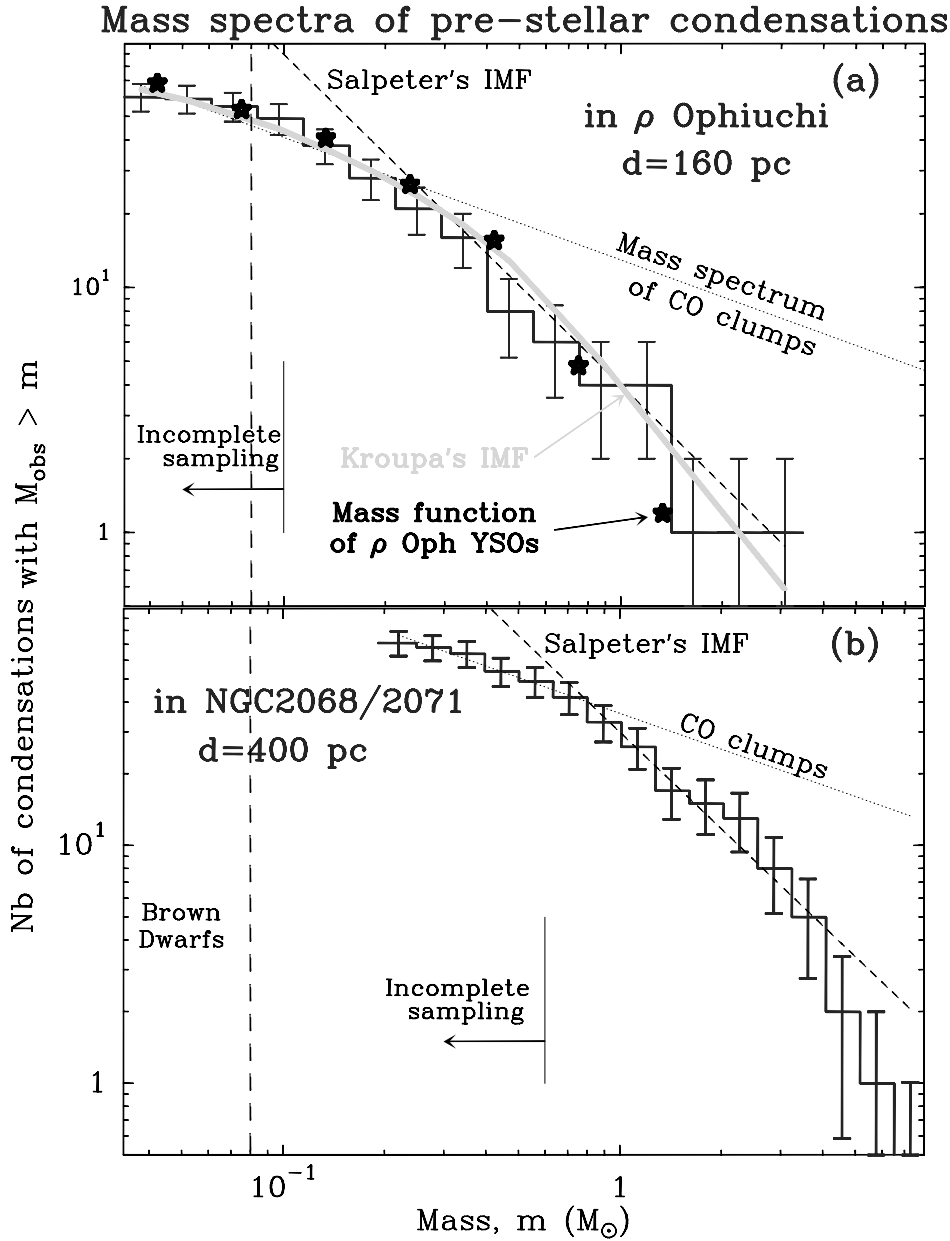}
\vspace{-0.3cm}
\caption{
Cumulative mass distributions of the pre-stellar condensations in the
$\rho$~Oph ({\bf a}) and NGC~2068/2071 (in {\bf b}) protoclusters
(from MAN98 and MAWB01).  The dotted and dashed lines show power-laws
of the form $N(>m) \propto m^{-0.5}$ (mass spectra of CO clumps, see
Williams et al. 2000) and $N(>m)\propto m^{-1.35}$ (IMF of Salpeter
1955), respectively.  The continuous curve in {\bf (a)} shows the
shape of the field star IMF (Kroupa et al. 1993).  The star markers in
{\bf (a)} represent the mass function of $\rho$~Oph YSOs derived from
an extensive mid-IR survey with ISOCAM (Bontemps et al. 2001; Kaas \&
Bontemps, this volume).  The error bars correspond to $\sqrt N$
counting statistics.}
\end{figure}

\begin{figure}
\plotfiddle{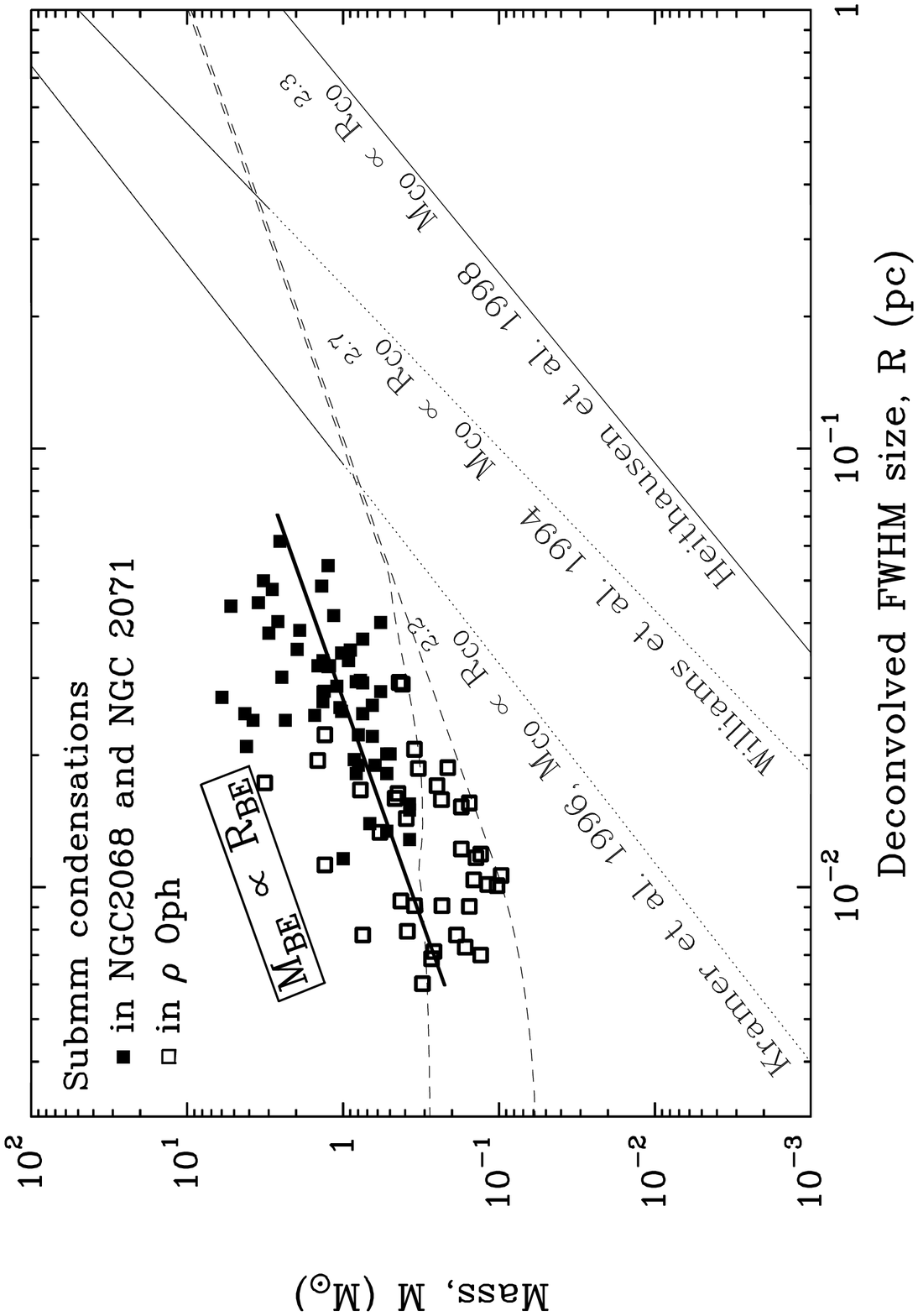}{7.1cm}{-90}{42}{42}{-180}{230}
\caption{
Mass-size relation of the starless submm condensations identified by
MAN98 and MAWB01 in the $\rho$~Oph and NGC~2068/2071 protoclusters
(open and filled squares, respectively). The thick solid line is the
mass-size relation expected for critical Bonnor-Ebert spheres (see
text). The two dashed curves show the 5$\sigma $ detection threshold
as a function of size in the two protoclusters.}
\end{figure}

\subsection{Mass-size relation of protocluster condensations}

Figure~5 compares the mass-size relations derived for the (sub)mm dust
continuum condensations of the $\rho$~Oph and NGC~2068/2071
protoclusters (from MAN98 and MAWB01) with those found for CO clumps
in various clouds (e.g. Heithausen et al.\ 1998).  The (sub)mm
condensations are definitely denser ($n_{\mbox{\tiny H2}} \ga
10^6~\mbox{cm}^{-3}$) than both CO clumps ($n_{\mbox{\tiny H2}} \sim
10^3~\mbox{cm}^{-3}$) and typical NH$_3$ cores ($n_{\mbox{\tiny H2}}
\sim 10^4-10^5~\mbox{cm}^{-3}$ -- Jijina, Myers, \& Adams 1999).  The
mass-size relation of the condensations spans only one decade in size
and is much flatter than that of CO clumps: a formal fitting analysis
gives $M_{\rm smm} \propto R_{\rm smm}\,^{1.1}$ as opposed to $M_{\rm
CO} \propto R_{\rm CO}\,^{2.4}$. Although the observed correlations
may be partly affected by size-dependent detection thresholds, it is
worth pointing out that they are suggestive of a change from a
turbulence-dominated to a gravity-dominated regime.  Indeed, while the
Larson law $M \propto R^2$ is consistent with the fractal, turbulent
nature of molecular clouds (e.g. Elmegreen \& Falgarone 1996), a
linear relation ($M \propto R$) is expected for a sample of
self-gravitating isothermal condensations assuming a uniform
temperature and a range of external pressures (cf. Bonnor 1956).  The
thick solid line in Fig.~5 displays the $M(R) = 2.4\, R \, a^2 /G $
relation obtained for critical Bonnor-Ebert spheres when $T = 15$~K
($a$ is the sound speed).  A very similar linear relation ($M(R) = 2\,
R \, a^2 /G $) holds for truncated singular isothermal spheres
(cf. Shu 1977). The largest condensations of NGC~2068/2071 are
consistent with $P_{\rm s} \sim 10^5 \ k_{\rm B}\ \mbox{cm}^{-3}\,$K,
and the smallest condensations of $\rho$~Oph with $P_{\rm s} \sim 10^7
\ k_{\rm B}\ \mbox{cm}^{-3}\,$K. These values are compatible with the
range of ambient pressures expected in protoclusters (cf. Myers 1998
and Johnstone et al. 2000).  Although the Bonnor-Ebert isothermal
model is probably too simplistic to fully describe the condensations
observed in protoclusters, the comparison shown in Fig.~5 is quite
encouraging.  It illustrates that {\em the pre-stellar condensations
identified in the dust continuum are much more centrally concentrated
than CO clumps and require the effects of self-gravity}.

\section{Conclusions}
  
There is now a growing body of evidence that the fragmentation of
dense ($\sim 10^5-10^6~{\rm cm}^{-3}$) cores into compact,
self-gravitating condensations determines the IMF of star clusters in
the low- to intermediate-mass range ($0.1-5\ M_\odot$).  A plausible
scenario, supported by some numerical simulations of cluster formation
(Klessen \& Burkert 2000; Padoan et al. 2001), could be the following:
First, cloud turbulence generates a field of density fluctuations, a
fraction of them corresponding to self-gravitating fragments; second,
these fragments (or ``kernels'') decouple from their turbulent
environment (e.g. Myers 1998) and collapse to protostars after little
interaction with their surroundings.

More extensive surveys should be done to improve the statistics and
search for starless condensations more massive than $10~M_\odot$.
However, high-mass stars may not form from the collapse of single
condensations but from the merging of several pre-/proto-stellar
condensations of low to intermediate mass.  In the collision scenario
of Bonnell, Bate, \& Zinnecker (1998), the cluster crossing time must
be short enough to allow individual condensations to collide and
coalesce with one another. Follow-up high-resolution dynamical studies
in dense molecular tracers are thus necessary to decide whether the
condensations detected in the submm dust continuum have the potential
to form massive stars with $M_\star \ga 10~M_\odot$. Preliminary
results of such a dynamical study suggest that this is not the case in
the $\rho$~Oph protocluster (Belloche et al., this volume).

Deeper, wider submillimeter continuum surveys in a variety of
star-forming regions are clearly required to set stronger constraints
on the IMF's origin and investigate possible environmental effects. At
the end of the present decade, the FIRST/Herschel satellite equipped
with the bolometer arrays SPIRE and PACS will carry out unbiased
surveys of nearby ($d \la 1$~kpc) cloud complexes at 90--300~$\mu$m,
i.e., at wavelengths where cold pre-stellar condensations emit most of
their radiation. Thanks to its unprecedented resolution and
sensitivity around $\lambda \sim $~0.8--1~mm (and below), the ALMA
interferometer will make possible similar investigations in distant
protoclusters including Galactic mini-starburst regions.

\acknowledgments
We acknowledge useful discussions with Anthony Whitworth on the
mass-size relation of submm condensations given in Fig.~5.

\end{document}